\documentclass[groupedaddress,
 reprint,
 amsmath,amssymb,
 nofootinbib,
 superscriptaddress,
 aps,
 prd,
]{revtex4-1}

\usepackage{graphicx}
\usepackage{dcolumn}
\usepackage{bm}
\usepackage{hyperref}
\usepackage[utf8]{inputenc}
\usepackage[capitalise]{cleveref}
\usepackage{aas_macros}
\usepackage[normalem]{ulem}
\usepackage{xcolor}
\usepackage{csquotes}

\usepackage[makeroom]{cancel}
\usepackage{printlen}

\usepackage{mathtools}
\usepackage{physics}

\usepackage{amsmath}
\makeatletter
\newcommand*\rel@kern[1]{\kern#1\dimexpr\macc@kerna}
\newcommand*\widebar[1]{%
  \begingroup
  \def\mathaccent##1##2{%
    \rel@kern{0.8}%
    \overline{\rel@kern{-0.8}\macc@nucleus\rel@kern{0.2}}%
    \rel@kern{-0.2}%
  }%
  \macc@depth\@ne
  \let\math@bgroup\@empty \let\math@egroup\macc@set@skewchar
  \mathsurround\z@ \frozen@everymath{\mathgroup\macc@group\relax}%
  \macc@set@skewchar\relax
  \let\mathaccentV\macc@nested@a
  \macc@nested@a\relax111{#1}%
  \endgroup
}
\makeatother
\newcommand{\mean}[1]{\widebar{#1}}

\begin{document}

\title{Stochastic Gravitational Waves from Post-inflationary Structure Formation}

\author{Benedikt Eggemeier}
\email{benedikt.eggemeier@phys.uni-goettingen.de}
\affiliation{
 Institut f\"ur Astrophysik, Georg-August-Universit\"at G\"ottingen, D-37077 G\"ottingen, Germany
}

\author{Jens C. Niemeyer}
\email{jens.niemeyer@phys.uni-goettingen.de}
\affiliation{
 Institut f\"ur Astrophysik, Georg-August-Universit\"at G\"ottingen, D-37077 G\"ottingen, Germany
}

\author{Karsten Jedamzik}
\email{karsten.jedamzik@umontpellier.fr}
\affiliation{Laboratoire Univers et Particules de Montpellier (LUPM),
Université de Montpellier (UMR-5299) CNRS, Place Eugène Bataillon
F-34095 Montpellier Cedex 05, France}

\author{Richard Easther}
\email{r.easther@auckland.ac.nz}
\affiliation{Department of Physics, University of Auckland, Private Bag 92019, Auckland, New Zealand}

\date{\today}

\begin{abstract}
Following inflation, the Universe may pass through an early  matter-dominated phase supported by the oscillating inflaton condensate. Initially small fluctuations in the condensate grow gravitationally on subhorizon scales and can collapse to form nonlinear ``inflaton halos''. Their formation and subsequent tidal interactions  will source gravitational waves, resulting in a stochastic background in the present Universe. We extend N-body simulations that model the growth and interaction of collapsed structures to compute the resulting  gravitational wave emission. The spectrum of this radiation is well-matched by semi-analytical estimates based on the collapse of inflaton halos and their tidal evolution. We  use this semi-analytic formalism to infer the spectrum for scenarios where the early matter-dominated phase gives way to a thermalized universe at temperatures as low as $100\,\mathrm{MeV}$ and we discuss the possible experimental opportunities created by this signal in inflationary models in which thermalization takes place long after inflation has completed. 
\end{abstract}

\maketitle

\section{Introduction}

Cosmological gravitational wave backgrounds propagate freely and thus provide direct information about the state of the universe at the time of their production. Low-frequency primordial gravitational waves can be detected via the B-mode of the polarization of the Cosmic Microwave Background. In contrast, higher-frequency gravitational waves can be observed via direct detection experiments.  

In contrast to the gravitational wave spectrum originating from the quantization of inflationary tensor perturbations~\cite{Abbot1984, Lucchin1985_1, Lucchin1985_2, Allen1988, Lyth1992, Stewart1993}, a gravitational wave background can be produced classically in scenarios where large and time-dependent density fluctuations are present in the early universe. A number of different processes  in the post-inflationary universe can generate such inhomogeneities. The resulting gravitational waves vary widely in strength and frequency as a function of the specific scenario and would be observed at the present time via a stochastic gravitational wave background~(SGWB). 

Many experiments have been proposed or are already in operation to search for gravitational waves at frequencies ranging from $10^{-9}\,\mathrm{Hz}$ to $10^3\,\mathrm{Hz}$. 
These include the future space-based interferometers LISA~\cite{LISA}, DECIGO~\cite{DECIGO}, BBO~\cite{BBO} and $\mu$Ares~\cite{muAres}, currently operating terrestrial interferometers (Advanced) LIGO~\cite{LIGO_1,LIGO_2}, Advanced Virgo~\cite{Virgo} and KAGRA~\cite{Kagra} and future proposals such as the Einstein telescope~\cite{ET}.  Pulsar timing arrays~\cite{PTAs} become sensitive to gravitational waves at even lower frequencies. 
It is thus crucial to predict the shape of the gravitational wave spectra from specific post-inflationary sources in order to assess their observability by present-day and future experiments.  

Inflationary cosmology proposes that the very early universe undergoes a period of accelerated expansion~\cite{Starobinsky1980, Guth1981, Linde1982, Linde1983}. In simple scenarios, this is driven by a scalar field, or inflaton, whose potential energy decreases slowly as the Universe expands.  Immediately after inflation ends the inflaton oscillates around the minimum of its potential. Depending on the explicit shape of the potential and the couplings to other fields, inflaton oscillations can resonantly amplify the occupation numbers of certain momentum modes of both coupled fields and their own fluctuations. 
These \textit{preheating} mechanisms~\cite{Traschen:1990sw, Shtanov:1994ce, Kofman1997} lead to strong fluctuations in density fields in the post-inflationary universe and the violent motion associated with resonance can source gravitational waves~\cite{Khlebnikov1997_grav, Easther2006_1, Easther2006_2, Garcia-Bellido2007_1, Easther2007, Garcia-Bellido2007_2, Dufaux2007, Dufaux2008, Figueroa2017}. Moreover, long-lived oscillons can form in certain parametric resonance scenarios~\cite{Gleiser:1993pt, Copeland:1995fq, Amin2010, Amin2012, Lozanov:2017hjm} which would be accompanied by a characteristic SGWB~\cite{Zhou2013, Antusch2016}.

Conversely, if the Universe does not pass through a resonant phase the  inflaton field oscillations will slowly damp, while the overall dynamics will resemble those of a matter-dominated universe.   In this case, the dominant inflaton interactions are gravitational, and initially small primordial fluctuations in the inflaton field grow once they are inside the Hubble horizon~\cite{Jedamzik:2010hq, Easther2010}. The momenta of typical quanta will be small relative to their rest mass, so the resulting dynamics of this self-gravitating quantum matter can be described by the non-relativistic Schrödinger-Poisson equations \cite{Musoke2019}. The initially small overdensities can easily grow to the point where they collapse to form gravitationally bound structures prior to thermalization~\cite{Musoke2019, Niemeyer2019, Eggemeier2020, Eggemeier2021}.

\begin{figure*}
    \centering
    \includegraphics[width=0.75\textwidth]{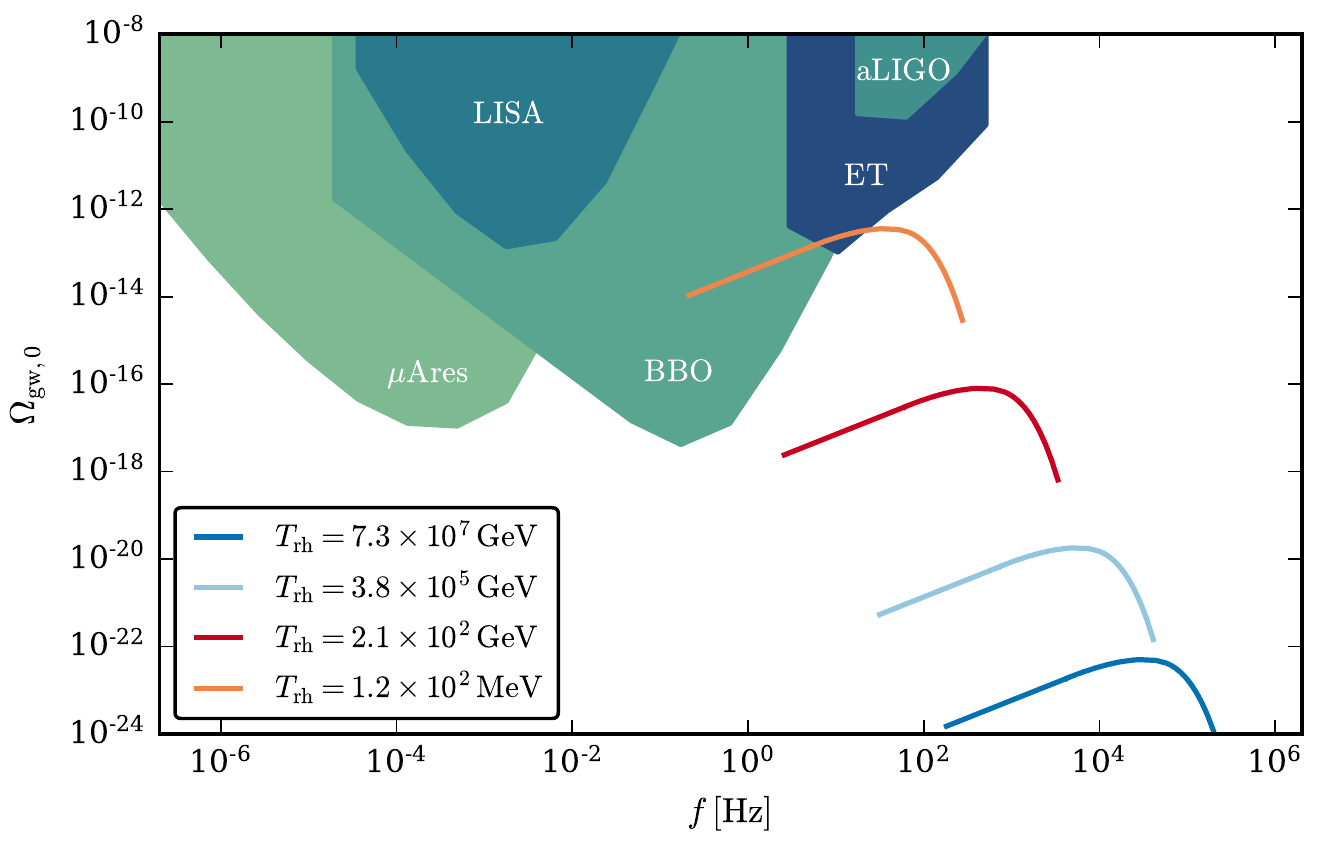}
    \caption{Stochastic gravitational wave background at the present time assuming different reheating temperatures. The sensitivity curves on the energy density of GWs for the experiments LISA, BBO, advanced LIGO, $\mu$Ares, and the Einstein telescope (ET) are shown for comparison. The data of the experimental limits were taken from Ref.~\cite{Campeti2020}.} 
    \label{fig:Omega_gw_experiments}
\end{figure*}

The Schrödinger-Poisson equations also govern the evolution of ultralight or fuzzy dark matter~(FDM)~\cite{Hu2000} in the late-time universe, so computational methods used in FDM-based cosmological structure formation simulations can be applied directly to this phase in the very early universe. This analogy was exploited in Ref.~\cite{Eggemeier2020} to perform large N-body simulations of the gravitational fragmentation of the inflaton field. These showed that the inflaton field  collapses into  inflaton halos with a mass distribution in agreement with the expectations from a Press-Schechter approach~\cite{Niemeyer2019}. The N-body simulations were then extended so that the Schrödinger-Poisson equations were directly solved at the finest levels of adaptively refined grids, revealing the formation of solitonic cores -- or inflaton stars -- in the center of inflaton halos~\cite{Eggemeier2021}. If the matter-dominated era lasts for 30 $e$-folds of growth the most massive inflaton stars might collapse to form primordial black holes (PBHs), a possibility further discussed in Refs.~\cite{Padilla2022, Hidalgo2022}. 

The dynamics of large, localized overdensities suggest that this early era of structure formation constitutes a further potential source of gravitational waves in the early universe.\footnote{Other mechanisms apart from the preheating scenarios mentioned earlier include first-order phase transitions~\cite{Kosowsky1992, Kaminokowski1994, Grojean2007}, networks of cosmic strings~\cite{Vilenkin1981} and the Hawking radiation of gravitons from a decaying population of small PBHs~\cite{Anantua2009}.} PBH formation is not the only potential gravitational wave source, as growing  density fluctuations in the early matter-dominated epoch will emit gravitational waves~\cite{Assadullahi2009, Jedamzik2010_grav}. Analytical estimates of the gravitational wave signal from the collapse of density fluctuations into inflaton halos, as well as the contribution of tidal effects afterwards and the evaporation of inflaton halos at reheating, were obtained in Ref.~\cite{Jedamzik2010_grav}. Based on previous simulations~\cite{Eggemeier2020, Eggemeier2021}, we complement the results from Ref.~\cite{Jedamzik2010_grav} by computing the SGWB from full numerical N-body simulations running from 14 to 23 $e$-folds after the end of inflation. We find that the numerical spectrum can be well described by a combination of the signal from spherical collapse and subsequent tidal interactions. This allows us to extrapolate the SGWB to reheating temperatures as low as $100\,\mathrm{MeV}$ where the signal lies within the sensitivity range of the  experimental proposals of BBO and the Einstein telescope, as illustrated in \cref{fig:Omega_gw_experiments}.

The structure of this paper is as follows. In \cref{sec:EMD_GWs} we briefly review the early matter-dominated epoch and provide an estimate for the gravitational wave signal from the collapse of density perturbations into inflaton halos. The setup of our N-body simulations and the implementation of the computation of the SGWB are described in \cref{sec:simulations}, followed by our numerical results. We discuss observational prospects in \cref{sec:discussion} and, finally,  we conclude in \cref{sec:conclusions}.

\section{Early Matter-dominated Epoch and generation of gravitational waves}
\label{sec:EMD_GWs}

Depending on the effective potential of the scalar inflaton field $\varphi$, inflation can be followed by an extended phase of matter-dominated expansion prior to reheating. Once slow-roll inflation has come to an end, the inflaton performs oscillations around the minimum of its potential. The inflaton potential $V(\varphi)$ is required to be shallower than quadratic at field values larger than the reduced Planck mass $M_\mathrm{Pl} = (8\pi G)^{-1/2}$ in order to be consistent with observations~\cite{Planck2018_inflation}. At field values smaller than $M_\mathrm{Pl}$ the inflaton potential can often be approximated by a quadratic potential around its minimum,
\begin{align}
    V(\varphi) = \frac{1}{2}m^2\varphi^2\,.
    \label{eq:quadratic_potential}
\end{align}
Assuming that the higher-order terms of the full potential do not support resonance, the inflaton field evolves as $\varphi (t) \sim\sin(mt)/t$ after the end of inflation. Averaged over several oscillations, the scale factor grows as $a(t)\sim t^{2/3}$ and the Hubble parameter reduces as $H\sim a^{-3/2}$~\cite{Albrecht:1982mp}. Thus, the post-inflationary evolution resembles expansion in a purely matter-dominated universe. 

This epoch continues as long as the Hubble parameter is larger than the decay rate $\Gamma$ of the inflaton. Provided the coupling of the inflaton to other fields is small, this era can last for multiple $e$-folds of expansion. When $H\simeq \Gamma$, the inflaton decays into radiation, a process known as reheating. The energy scale of reheating is usually given in terms of the temperature~\cite{Kofman1997}
\begin{align}
    T_\mathrm{rh} \simeq 0.55\left(\frac{100}{g_\ast}\right)^{1/4}(\Gamma M_\mathrm{Pl})^{1/2}\,,
    \label{eq:reheating_temp}
\end{align}
where $g_\ast$ is the number of relativistic degrees of freedom at reheating. 

During the matter-dominated era, subhorizon density perturbations grow gravitationally and finally collapse to form inflaton halos and inflaton stars~\cite{Jedamzik:2010hq, Easther2010, Musoke2019, Niemeyer2019, Eggemeier2020, Eggemeier2021}. It was shown in Ref.~\cite{Jedamzik2010_grav} that this early phase of gravitational structure formation is associated with the generation of gravitational waves. They can be sourced, for example, by the formation and the subsequent tidal interactions of inflaton halos. In the following, we will estimate the gravitational wave signal from the formation of inflaton halos. 

Gravitational waves are represented by the spatial tensor perturbations $h_{ij}$ of the  FLRW metric, which can be written to leading order as~\cite{Caprini2018}
\begin{align}
    \mathrm{d}s^2 = -\mathrm{d}t^2 +  a^2(t) (\delta_{ij} + h_{ij})\mathrm{d}x^i\mathrm{d}x^j\,.
\end{align}
In this gauge, the tensor perturbations satisfy the conditions $\partial_i h_{ij} = h_{ii} = 0$ and are thus symmetric, transverse and traceless. Switching to conformal time $\mathrm{d}\tau = \mathrm{d}t/a$, their evolution is governed by~\cite{Caprini2018}
\begin{align}
    h_{ij}^{\prime\prime} + 2\mathcal{H}h_{ij}^\prime - \nabla^2 h_{ij} = 16\pi G a^2 \Pi_{ij}^\mathrm{TT}\,,
    \label{eq:GW_eom}
\end{align}
where a prime denotes a derivative with respect to conformal time, $\mathcal{H} = aH$, and $\Pi_{ij}^\mathrm{TT}$ is the traceless and transverse part of the anisotropic stress tensor. The full anisotropic stress tensor $\Pi_{ij}$ is given by
\begin{align}
    a^2\Pi_{ij} = T_{ij} - \mean{p}a^2(\delta_{ij}+h_{ij})\,,
    \label{eq:anisotropic_stress_T}
\end{align}
where $T_{ij}$ denotes the spatial components of the stress-energy tensor of the 
inflaton field and $\mean{p}$ is the background pressure. Since the $\mean{p}\delta_{ij}$ term in \cref{eq:anisotropic_stress_T} is a pure trace, it does not contribute to $\Pi_{ij}^\mathrm{TT}$. The metric perturbation $h_{ij}$ appearing in the second term on the right-hand side of \cref{eq:anisotropic_stress_T} is subdominant on subhorizon scales and can be neglected~\cite{Dufaux2007}. 

During inflation the inflaton field is homogeneous and the source term on the right-hand side of \cref{eq:GW_eom} vanishes.
This is not the case in the succeeding matter-dominated era where the gravitational fragmentation of the inflaton field results in the generation of gravitational waves. One can estimate the radiated energy density using the quadrupole formula where the amplitude $h$ of the gravitational wave can be approximated as~\cite{Schutz1984, Jedamzik2010_grav}
\begin{align}
    h \simeq \frac{G}{2}\left(\Ddot{I}_{ij} - \frac{1}{3}\Ddot{I}_{kk}\delta_{ij}\right)\frac{n_i n_j}{|\mathbf{x}|}\,.
\end{align}
Here, $\mathbf{n}$ is the radial unit vector from the origin located in the center of the source to the point $\mathbf{x}$. The two terms in the brackets are the second time derivative of the traceless part of the quadrupole tensor $I_{ij}$. Assuming the formation of an inflaton halo with mass $M_h$, radius $R_h$ and virial velocity $v_h = (GM_h/R_h)^{1/2}$, an order-of-magnitude estimate for $\Ddot{I}_{ij}$ is~\cite{Schutz1984}
\begin{align}
    \Ddot{I}_{ij} = \frac{\mathrm{d}^2}{\mathrm{d}t^2}\int \rho x^2\,\mathrm{d}^3x \sim 2 \int\rho v_h^2\,\mathrm{d}^3x \sim 2M_h v_h^2\,,
\end{align}
which yields $h \sim GM_h v_h^2 n_i n_j/|\mathbf{x}|$ for the gravitational wave amplitude. Note that this should be understood as an upper limit since spherically symmetric motions do not generate gravitational waves, i.e. the formation of a perfectly spherical inflaton halo does not generate gravitational radiation. 

The energy flux of the gravitational wave is proportional to $\Dot{h}^2$ and the related luminosity of the source is~\cite{Schutz1984}
\begin{align}
    L_\mathrm{gw}^\mathrm{coll} \simeq \frac{|\mathbf{x}| \Dot{h}^2}{G} \sim \frac{\omega^2|\mathbf{x}|^2h^2}{G}\,,
\end{align}
where the gravitational wave frequency $\omega$ can be approximated by the natural dynamical frequency, given by the inverse of the free-fall time $t_\mathrm{coll} = (3\pi/(16 G\mean{\rho}))^{1/2}$  of the collapsing object. Thus, the luminosity of the source can be estimated as
\begin{align}
    L_\mathrm{gw}^\mathrm{coll} \sim \frac{G^4 M_h^5}{R_h^5 \pi^2}\,,
\end{align}
and the radiated gravitational wave energy is $E_\mathrm{gw} = L_\mathrm{gw}^\mathrm{coll} t_\mathrm{coll}$. 

In order to compute the gravitational wave signal not only from the collapse of a single halo but from the collapse of halos on separate length scales at different times during the  early structure formation phase, we assume that a perturbation on the (comoving) scale $k$ forms a halo once it has become nonlinear. 
This means that a halo of size $R_h = 2\pi a_\mathrm{coll}/k$ and mass $M_h = 4\pi/3 \mean{\rho}(a_\mathrm{coll})R_h^3$ radiates gravitational waves at a well-defined time and at frequency $f = k a_\mathrm{coll}/(2\pi)$, where $a_\mathrm{coll}$ denotes the scale factor at time of collapse. Taking into account that the energy in gravitational waves scales as $a^{-4}$, the energy density of the gravitational waves at the present time over a wide range of $k$ can be written as~\cite{Jedamzik2010_grav}
\begin{align}
     \frac{\mathrm{d}\rho_\mathrm{gw}^{\mathrm{coll},0}}{\mathrm{d}\ln k} &= \frac{\mathrm{d}n_h}{\mathrm{d}\ln k} L_\mathrm{gw}^\mathrm{coll} t_\mathrm{coll}\left(\frac{a_\mathrm{coll}}{a_0}\right)^4 \nonumber\\ 
     &= \frac{3\bar\rho(a_\mathrm{coll})}{M}L_\mathrm{gw}^\mathrm{coll}  t_\mathrm{coll}\left(\frac{a_\mathrm{coll}}{a_0}\right)^4\, ,
     \label{eq:rho_gw_spherical}
\end{align}
with $a_0$ the present-day scale factor. To obtain this expression we estimated the halo number density $n_h$ as $n_h = N_h/V$, where $N_h$ denotes the total number of halos, and the overall volume they cover is $V = N_h M/\mean\rho(a_\mathrm{coll})$. Hence $\mathrm{d}n_h/\mathrm{d}\ln M = -\mean{\rho}(a_\mathrm{coll})/M$ and an additional factor of $-3$ comes from considering $\mathrm{d}\ln k$ instead of $\mathrm{d}\ln M$. 

Assuming that thermal equilibrium is established at the reheating temperature $T_\mathrm{rh}$, the Hubble parameter at reheating is~\cite{Kofman1997} 
\begin{align}
    H_\mathrm{rh}^2 = \frac{\rho_\mathrm{rh}}{3M^2_\mathrm{Pl}} = \frac{g_\ast\pi^2 T_\mathrm{rh}^4}{90M^2_\mathrm{Pl}}\,.
    \label{eq:Hubble_reh}
\end{align}
Since $g_\ast a^3 T^3 = \mathrm{const}$ in thermal equilibrium we can express the scale factor dependence in \cref{eq:rho_gw_spherical} as $a_\mathrm{coll}/a_0 = (a_\mathrm{coll}/a_\mathrm{rh})(a_\mathrm{rh}/a_0)$. The expansion factor from $T_\mathrm{rh}$ to $T_0 = 2.7\,\mathrm{K} = 2.3\times 10^{-4}\,\mathrm{eV}$ is~\cite{Khlebnikov1997_grav}
\begin{align}
    \frac{a_\mathrm{rh}}{a_0} = \left(\frac{g_0}{g_\ast}\right)^{1/3}\frac{T_0}{T_\mathrm{rh}} = \frac{g_0^{1/3}}{g_\ast^{1/12}}\left(\frac{\pi^2}{90}\right)^{1/4}\frac{T_0}{(M_\mathrm{Pl}H_\mathrm{rh})^{1/2}}\,,
    \label{eq:scalefactor_reh_0}
\end{align}
where \cref{eq:Hubble_reh} was used to replace $T_\mathrm{rh}$. We will make use of \cref{eq:rho_gw_spherical} to compare the analytical prediction with the gravitational wave signal obtained directly from the structure formation simulations presented in the next section.

\section{Simulations of Early Structure Formation}
\label{sec:simulations}

We make use of \textsc{AxioNyx}~\cite{Schwabe2020}, which is based on the cosmology code \textsc{Nyx}~\cite{Almgren2013}, to perform N-body simulations of gravitational structure formation in the post-inflationary universe. Complementing previous simulations~\cite{Eggemeier2020, Eggemeier2021}, we extend \textsc{AxioNyx} to compute the SGWB that is associated with the formation of inflaton halos and their subsequent tidal interactions. Assuming chaotic inflation, i.e. the potential in \cref{eq:quadratic_potential}, the first halos form $\mathcal{N}\simeq 16.7$ $e$-folds after the end of inflation, so we start the computation of the gravitational wave signal $\mathcal{N}=14$ $e$-folds after the end of inflation. To include the generation of gravitational waves from as many sources as possible, the  simulations are evolved to $\mathcal{N}=23$ which corresponds to a reheating temperature of $T_\mathrm{rh} = 7.3\times 10^7\,\mathrm{GeV}$ (cf. \cref{eq:reheating_temp}).

\subsection{Initial conditions and simulation setup}

Building on the simulations from Refs.~\cite{Eggemeier2020, Eggemeier2021}, it is convenient to work again with the purely quadratic inflaton potential of \cref{eq:quadratic_potential} with  $m=6.35\times 10^{-6}\,M_\mathrm{Pl}$,  $\varphi_\mathrm{end} \approx M_\mathrm{Pl}$ and $H_\mathrm{end}\approx m/\sqrt{6}$ at the end of inflation. 
Given our intention to run the N-body simulations from $\mathcal{N}=14$ to $\mathcal{N}=23$ $e$-folds after the end of inflation we slightly adjust the unit system relative to the one chosen in Refs.~\cite{Eggemeier2020, Eggemeier2021}.
The comoving length unit is $l_u = e^{23} H_\mathrm{end}^{-1} = 3.04\times 10^{-19}\,\mathrm{m}$ and we chose a mass unit of $m_u = 10^{-13}\,\mathrm{kg}$. For the time unit $t_u = 6.50\times 10^{-22}\,\mathrm{s}$ the gravitational constant is $G = 10^{-10}\,l_u^3/(m_u t_u^2)$. In this unit system, the Hubble parameter at  $\mathcal{N}=23$ is $H_{23} = 6.49\,t_u^{-1}$ and the corresponding mean density is $\mean{\rho}_{23} = 5.02\times 10^{10}\,m_u/l_u^3$.   

We use the same initial density power spectrum as in Ref.~\cite{Eggemeier2020}, illustrated in Fig.~2 of that paper. The power spectrum is resolved with a box side length of $L=1200\,l_u$ and $256^3$ particles. This choice  allows us to run the simulations to $\mathcal{N}=23$ $e$-folds after the end of inflation without the $k$-scale corresponding to the size of the simulation box becoming nonlinear. Larger box sizes would allow simulations that run to later times but at the cost of  increasing particle numbers and grid size to ensure that the peak of the initial power spectrum is resolved. This  computational expense is  unnecessary, however, as we will show in \cref{sec:numerical_results}.

\subsection{Computation of SGWB}
\label{sec:computation_SGWB}

We compute the SGWB generated from the formation of inflaton halos and their tidal interactions using the numerical approach from Ref.~\cite{Dufaux2007}, originally developed for preheating simulations. Starting from the gravitational wave evolution \cref{eq:GW_eom}, we work in Fourier space with the convention that
\begin{align}
        T_{ij}(\mathbf{k}) = \int \frac{\mathrm{d}^3\mathbf{x}}{(2\pi)^{3/2}}T_{ij}(\mathbf{x}) e^{i\mathbf{k}\mathbf{x}}\,.
\end{align}
Introducing the variable $\widetilde{h}_{ij} = ah_{ij}$, \cref{eq:GW_eom} can be written in Fourier space as
\begin{align}
    \widetilde{h}_{ij}^{\prime\prime}(\mathbf{k}) + \left(k^2 - \frac{a^{\prime\prime}}{a}\right)\widetilde{h}_{ij}(\mathbf{k}) = 16\pi GaT_{ij}^\mathrm{TT}(\mathbf{k})\,.
    \label{eq:eom_grav_waves_TT}
\end{align}
The transverse traceless part of $T_{ij}(\mathbf{k})$ is obtained from
\begin{align}
    T_{ij}^\mathrm{TT}(\mathbf{k}) = \mathcal{O}_{ijlm}(\mathbf{k}) T_{lm}(\mathbf{k})\,,
    \label{eq:stress_energy_TT}
\end{align}
where $\mathcal{O}_{ijlm}(\mathbf{k})$ denotes the projection operator~\cite{Dufaux2007}
\begin{align}
    \mathcal{O}_{ijlm}(\mathbf{k}) = P_{il}(\mathbf{e}_\mathbf{k}) P_{jm}(\mathbf{e}_\mathbf{k}) - \frac{1}{2} P_{ij}(\mathbf{e}_\mathbf{k}) P_{lm}(\mathbf{e}_\mathbf{k})\,,
\end{align}
with
\begin{align}
    P_{ij}(\mathbf{e}_\mathbf{k}) = \delta_{ij} - e_{\mathbf{k},i} e_{\mathbf{k},j}\,.
\end{align}
Here, $\mathbf{e}_\mathbf{k} = \mathbf{k}/k$ is the unit vector in $\mathbf{k}$-direction. In a matter-dominated universe the $a^{\prime\prime}/a$  term in \cref{eq:eom_grav_waves_TT} behaves as $a^{\prime\prime}/a\sim a^2H^2$ and since $a^2H^2\ll k^2$ on subhorizon scales we omit this term. Assuming that gravitational waves are sourced between the initial ($\tau_i$) and final time ($\tau_f$) of the simulations this reduced equation of motion has the solution~\cite{Dufaux2007, Jedamzik2010_grav}
\begin{align}
    \widetilde{h}_{ij}(\mathbf{k}) = A_{ij}(\mathbf{k})\sin[k(\tau - \tau_f)] + B_{ij}(\mathbf{k})\cos[k(\tau - \tau_f)]\,,
\end{align}
where 
\begin{align}
    A_{ij}(\mathbf{k}) &= \frac{16\pi G}{k} \int_{\tau_i}^{\tau_f}\mathrm{d}\tau \cos[k(\tau_f - \tau)]a(\tau)T_{ij}^\mathrm{TT}(\tau,\mathbf{k})\,, \\
    B_{ij}(\mathbf{k}) &= \frac{16\pi G}{k} \int_{\tau_i}^{\tau_f}\mathrm{d}\tau \sin[k(\tau_f - \tau)]a(\tau)T_{ij}^\mathrm{TT}(\tau,\mathbf{k})\,.
\end{align}
Defining the energy density $\rho_\mathrm{gw}$ of the generated gravitational waves as an average over the simulation volume $V=L^3$, 
\begin{align}
     \rho_\mathrm{gw} = \frac{1}{32\pi Ga^4}\frac{1}{V}\int \mathrm{d}^3k \widetilde{h}_{ij}^\prime(\tau,\mathbf{k}) \widetilde{h}_{ij}^{\prime\ast}(\tau,\mathbf{k})\,,
\end{align}
and inserting the derivative of $\widetilde{h}_{ij}(\mathbf{k})$, the gravitational wave energy density is~\cite{Dufaux2007}
\begin{align}
    \rho_\mathrm{gw} &= \frac{4\pi G}{Va^4}\int \mathrm{d}^3k \sum_{i,j} \Bigg[\Bigg.  \abs{\int_{\tau_i}^{\tau_f}\mathrm{d}\tau\cos(k\tau) a(\tau) T_{ij}^\mathrm{TT}(\tau,\mathbf{k})}^2 
    \hspace*{-0.18cm} \nonumber\\
    &\hspace*{1.5cm} + \abs{\int_{\tau_i}^{\tau_f}\mathrm{d}\tau\sin(k\tau) a(\tau) T_{ij}^\mathrm{TT}(\tau,\mathbf{k})}^2 \Bigg.\Bigg]\,.
    \label{eq:energy_density_grav_waves_final}
\end{align}
Expressing $\mathrm{d}^3k = k^3\mathrm{d}(\ln k) \mathrm{d}\Omega_k$ in terms of the solid angle $\Omega_k$ in Fourier space and considering instead the energy density per logarithmic $k$-interval~\cite{Dufaux2007}
\begin{align}
    \left(\frac{\mathrm{d}\rho_\mathrm{gw}}{\mathrm{d}\ln k}\right)_{\tau > \tau_f} = \frac{S_k(\tau_f)}{a^4(\tau)}\,,
    \label{eq:grav_energydens_Sk}
\end{align}
provides a computationally efficient method of obtaining the gravitational wave signal. Instead of solving the full three-dimensional $k$-integral at different times in the simulation, it is convenient to compute the quantity
\begin{widetext}
\begin{align}
    S_k(\tau_f) = \frac{4\pi Gk^3}{V}\int\mathrm{d}\Omega_k\sum_{i,j}  \Bigg[\Bigg.\abs{\int_{\tau_i}^{\tau_f}\mathrm{d}\tau\cos(k\tau) a(\tau) T_{ij}^\mathrm{TT}(\tau,\mathbf{k})}^2 
     + \abs{\int_{\tau_i}^{\tau_f}\mathrm{d}\tau\sin(k\tau) a(\tau) T_{ij}^\mathrm{TT}(\tau,\mathbf{k})}^2 \Bigg.\Bigg]\,.
    \label{eq:grav_waves_Sk}
\end{align}
\end{widetext}
This is independent of the subsequent cosmological evolution and we can easily evaluate $S_k$ along different $k$-directions in our N-body simulations.
We define the SGWB at a certain time by dividing \cref{eq:grav_energydens_Sk} by the current critical density $\rho_c$. Assuming that reheating occurs at the end of our simulations, i.e. $a_f = a_\mathrm{rh}$, the fractional contribution of the SGWB per logarithmic wave vector interval to the critical density at the present time is
\begin{align}
\Omega_{\mathrm{gw},0}(k) =
\frac{1}{\rho_{c,0}}S_k(\tau_f)\left(\frac{a_\mathrm{rh}}{a_0}\right)^4\,,
    \label{eq:Omega_gw_present}
\end{align}
where $a_\mathrm{rh}/a_0$ is given by \cref{eq:scalefactor_reh_0}. Furthermore, the comoving wave numbers $k$ need to be converted to the physical frequencies at the present time, $f_0 = k a_\mathrm{rh}/(2\pi a_0)$. 

The crucial steps to computing $S_k$ in our N-body simulations are the determination of the transverse and traceless stress-energy tensor $T_{ij}^\mathrm{TT}$ at each time step and the numerical integration of the oscillatory time integrals in \cref{eq:grav_waves_Sk}. 
When the inflaton oscillates around the bottom of its potential the universe is effectively matter-dominated, so we can approximate the stress-energy tensor as that of a perfect fluid with vanishing pressure~\cite{Jedamzik2010_grav}. 
With this simplification $T_{ij}$ can be expressed in terms of the comoving matter density field $\rho$ and the peculiar velocity field $\mathbf{v}$ as $T_{ij} = \rho \mathbf{v}_i\mathbf{v}_j$. These quantities can be directly obtained from the particle information of our N-body simulations. This allows us to determine $T_{ij}$ at each time step; each component of $T_{ij}$ is then transformed to Fourier space and  \cref{eq:stress_energy_TT} is then used to yield $T_{ij}^\mathrm{TT}$. This requires a specific direction in Fourier space  for the projection; following Ref.~\cite{Dufaux2007} we consider unit vectors in the six different $k$-directions: $(1,1,0)$, $(1,0,1)$, $(0,1,1)$, $(-1,1,0)$, $(-1,0,1)$, $(0,-1,1)$.

At each step, the two time integrals in \cref{eq:grav_waves_Sk} are summed and $S_k$ is computed along each of the chosen $k$-directions. The integral is performed in Fourier space and we need to resolve the temporal frequency of the smallest  scales in the simulation volume. This requires much smaller time steps than the underlying N-body simulation, increasing the number of steps from $\mathcal{N}=14$ to $\mathcal{N}=23$ by a factor of $\sim 500$.

\subsection{Numerical Results}
\label{sec:numerical_results}

\begin{figure}
    \centering
    \includegraphics[width=\columnwidth]{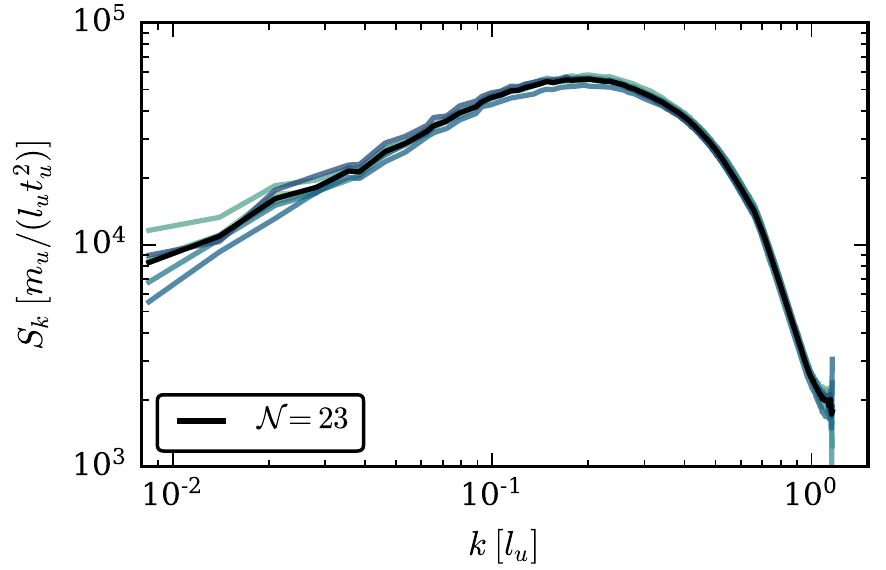}
    \caption{Spectrum $S_k$,  $\mathcal{N}=23$ $e$-folds after the end of inflation. The box side length is $L=1200\,l_u$ and the grid size is $256^3$. The six lines correspond to the individual $k$-directions; the black line shows the averaged spectrum.}
    \label{fig:Sk_N23}
\end{figure}

The spectra found for the six different $k$-directions at $\mathcal{N}=23$ are shown in \cref{fig:Sk_N23}. Evidently, the gravitational wave background is isotropic. In what follows we will show the spectrum obtained from averaging over all six specified directions in $k$-space. It is illustrated by the black curve in \cref{fig:Sk_N23}. 

\begin{figure*}
    \centering
    \includegraphics[width=\textwidth]{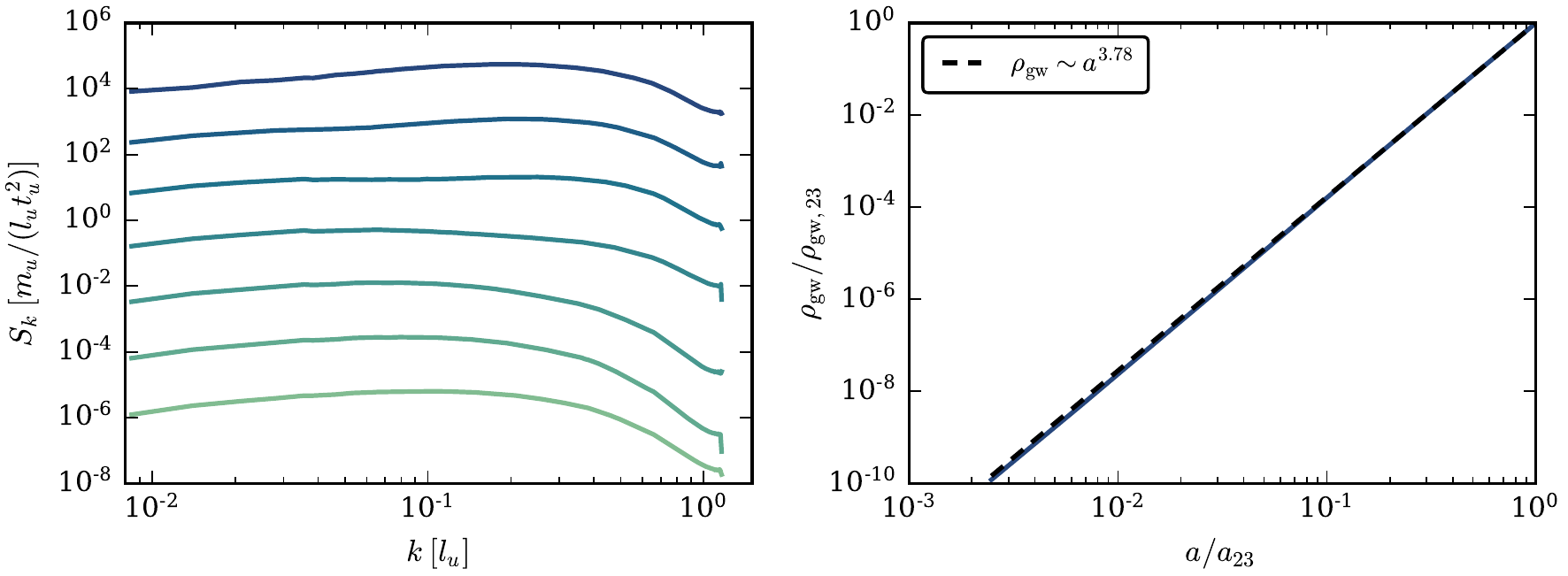}
    \caption{Left: evolution of $S_k$ from $\mathcal{N}=17$ (light green curve) to $\mathcal{N}=23$ $e$-folds (dark blue curve) after the end of inflation. Right: energy density of gravitational waves as a function of the scale factor. The black dashed line shows the power-law fit $\rho_\mathrm{gw}\sim a^{3.78}$.}
    \label{fig:Sk_evolution}
\end{figure*}

The evolution of $S_k$ from $\mathcal{N}=17$ to $\mathcal{N}=23$ is shown in the left panel of \cref{fig:Sk_evolution}. We observe an overall increase in power on all scales. While the shape of the spectrum on large scales is subject to only minor variations, a more pronounced alteration is visible on the smaller scales. Thanks to the comparatively large number of small inflaton halos and their formation at early times, the spectrum at scales $k \gtrsim 0.1\,l_u^{-1}$ is dominated by merger events and tidal interactions. The  gravitational wave signal from larger scales is less pronounced as the corresponding inflaton halos form later, are less dense, and have a lower frequency of tidal interactions.
The integral
\begin{align}
    \rho_\mathrm{gw}(\tau) = \int S_k(\tau)\,\mathrm{d}\ln k
\end{align}
gives the comoving energy density contained in gravitational waves in our simulation as a function of time. The evolution of $\rho_\mathrm{gw}$ as a function of scale factor can be seen on the right-hand side of \cref{fig:Sk_evolution} and is well described by the single power-law $\rho_\mathrm{gw}\sim a^{3.78}$. This allows us to extrapolate the numerically obtained gravitational wave spectrum to later times. 

\begin{figure*}
    \centering
    \includegraphics[width=\textwidth]{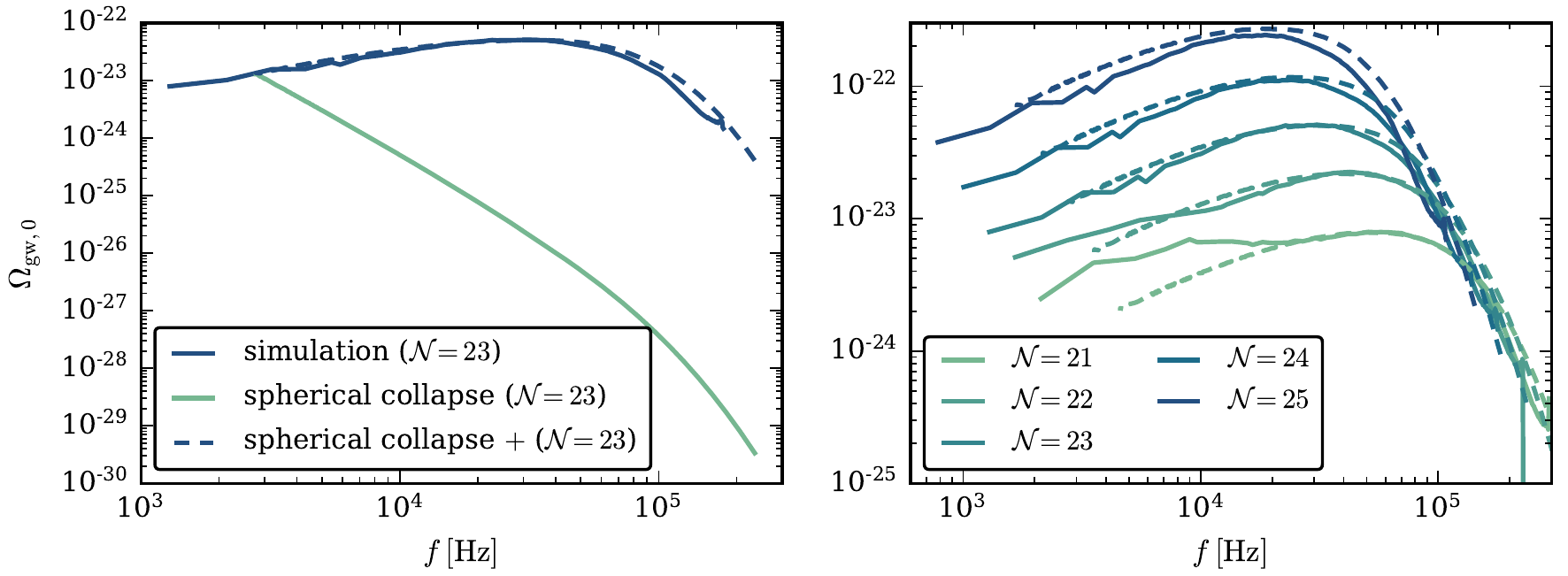}
    \caption{Comparison between the numerically obtained SGWB at the present time and the SGWB from the spherical collapse model. Left: spectra assuming reheating at $\mathcal{N}=23$ $e$-folds after the end of inflation from simulation, spherical collapse (see \cref{eq:rho_gw_spherical}) and spherical collapse complemented by further gravitational wave emission (see \cref{eq:rho_gw_spherical_mergers}) with parameters $p=1.84$ and $\xi=0.4$. Right: numerical spectra (solid lines) and spherical collapse complemented by further gravitational wave emission (see \cref{eq:rho_gw_spherical_mergers}) with unchanged parameters $p$ and $\xi$ (dashed lines) assuming reheating at the respective times. The simulation results from $\mathcal{N}=23$ were extrapolated to later times by taking into account an overall power increase of $\sim a^{0.78}$ (see text for details).}
    \label{fig:Omega_gw0_reh}
\end{figure*}

Assuming that reheating takes place at the end of our simulations, i.e. at the reheating temperature $T_\mathrm{rh} = 7.3\times 10^7\,\mathrm{GeV}$, we use \cref{eq:Omega_gw_present} to obtain the SGWB 
at the present time. As is visible from \cref{fig:Omega_gw0_reh}, the spectrum peaks at a frequency of $3\times 10^4\,\mathrm{Hz}$ with an amplitude of $\Omega_{\mathrm{gw},0}^\mathrm{max} \simeq 5\times 10^{-23}$. 

We can compare the numerical SGWB with the signal associated with the singular emission of gravitational waves from the spherical collapse of an inflaton halo on a  scale $k$ (see \cref{eq:rho_gw_spherical}). Evolving the initial power spectrum linearly with the growth factor, it starts to deviate from the numerical density power spectrum once it crosses the threshold $\Delta^2(k)\simeq 0.1$. We can thus expect that once a  scale crosses this threshold it will collapse into a halo. The radius of such a collapsing object is  given by $R_h = 2\pi a_\mathrm{coll}/k$ and  its mass is $M_h = 4\pi/3 \mean\rho(a_\mathrm{coll})R_h^3$. Using \cref{eq:rho_gw_spherical} and considering halos forming up to $\mathcal{N}=23$, we  compute the resulting gravitational wave signal from spherical collapse at the present time. This is shown in the left panel of \cref{fig:Omega_gw0_reh}. 

Since the gravitational waves originating from the early collapse of small-scale halos are highly redshifted, the strongest contribution comes from high-mass objects that collapse at late times. This appears to differ from the SGWB obtained numerically -- that is a flatter spectrum with strong contributions on small scales. The signal from spherical collapse is smaller by several orders of magnitude at high frequencies, pointing to a significant contribution originating from mergers and other tidal interactions on small scales, as these are only captured by the full N-body solution.  

Following Ref.~\cite{Jedamzik2010_grav}, we extend \cref{eq:rho_gw_spherical} to take into account
the missing contributions from the post-virialization rotation/vibration of halos and the occasional tidal interactions among each other. 
Between the end of spherical collapse and reheating, the physical background density of halos decreases as $\rho\sim a^{-3}$  introducing an extra factor of $(a_\mathrm{coll}/a)^3$ in Eq.~(\ref{eq:rho_gw_spherical})for  signals emitted after spherical collapse. On the other hand, the emission may occur well after collapse, i.e. $t_\mathrm{coll} \rightarrow t_\mathrm{coll}(a/a_\mathrm{coll})^{3/2}$.
Finally, with later emission the redshift of the gravitational wave energy density is much reduced, enhancing the signal by $(a/a_\mathrm{coll})^{4}$. 
Assuming a reduced efficiency $\xi < 1$ for post-collapse emission, the resulting signal can be estimated as~\cite{Jedamzik2010_grav}
\begin{align}
    \frac{\mathrm{d}\rho_\mathrm{gw}^{>\mathrm{coll},0}}{\mathrm{d}\ln k} &= \xi \frac{3\bar\rho(a_\mathrm{coll})}{M}  \left(\frac{a_\mathrm{coll}}{a_\mathrm{rh}}\right)^3 Lt_\mathrm{coll}  \left(\frac{a_\mathrm{rh}}{a_\mathrm{coll}}\right)^{3/2}\left(\frac{a_\mathrm{rh}}{a_0}\right)^4\nonumber\\
    &= \xi\frac{\mathrm{d}\rho_\mathrm{gw}^{\mathrm{coll},0}}{\mathrm{d}\ln k} \left(\frac{a_\mathrm{rh}}{a_\mathrm{coll}}\right)^{5/2}\,.
\end{align}
Instead of taking the value of the exponent at face value, we treat it as a free parameter $p$ such that
\begin{align}
    \frac{\mathrm{d}\rho_\mathrm{gw}^{>\mathrm{coll},0}}{\mathrm{d}\ln k} = \xi\frac{\mathrm{d}\rho_\mathrm{gw}^{\mathrm{coll},0}}{\mathrm{d}\ln k}  \left(\frac{a_\mathrm{rh}}{a_\mathrm{coll}}\right)^{p}\,,
    \label{eq:rho_gw_spherical_mergers}
\end{align}
and fit $p$ together with the efficiency parameter $\xi$ to the numerical  spectrum. The blue dashed curve in \cref{fig:Omega_gw0_reh} corresponds to the fitted parameters $p=1.84$ and $\xi=0.4$ and accurately describes the numerical spectrum at $\mathcal{N}=23$. It is not surprising that we find $p< 5/2$, as one can imagine that over time tidal interactions and irregularities in halos reduce. It is noteworthy that the gravitational wave spectra at other times are reasonably well approximated for fixed  $p$ and $\xi$,  as seen in the right panel of \cref{fig:Omega_gw0_reh}. 

Note that the final simulation spectrum at $\mathcal{N}=23$ was extrapolated to $\mathcal{N}=24$ and $\mathcal{N}=25$, respectively, making use of our previous result from \cref{fig:Sk_evolution} that the energy density of gravitational waves increases with $a^{3.78}$. 
Additionally, the $a_\mathrm{rh}^4$ dependence in  \cref{eq:Omega_gw_present} leads to an increase of $a^3$ in the spectrum $\Omega_{\mathrm{gw},0}$. At the same time, we also need to take into account that the physical matter density field, which is involved in the computation of $T_{ij}^\mathrm{TT}$ and enters \cref{eq:grav_waves_Sk} quadratically, decreases by $a^3$. Combining the single contributions, one obtains an overall increase of $a^{0.78}$ in $\Omega_{\mathrm{gw},0}$ when reheating takes place at later times.

It is possible that the calibration of $\xi$ depends on the specific  inflationary model. Also, increasing the spatial resolution of the simulation could lead to slightly different parameter choices as better resolution of small-scale objects would reveal more gravitational wave sources at high frequencies. 
However, this does not affect the signal on large scales. 

\section{Observational Prospects}
\label{sec:discussion}

Having calibrated \cref{eq:rho_gw_spherical_mergers} to our simulations we can extrapolate it to give the SGWB at later times. \cref{fig:Omega_gw_experiments} shows the evolution of the spectrum from $\mathcal{N}=23$ to $\mathcal{N}=50$, corresponding to reheating temperatures of $T_\mathrm{rh} = 7.3\times 10^7\,\mathrm{GeV}$ and $T_\mathrm{rh} = 1.2\times 10^2\,\mathrm{MeV}$ respectively. The latter, very low, temperature is the absolute minimum value consistent with nucleosynthesis~\cite{Kawasaki1999, Hannestad2004, Salas2015}. We show the expected sensitivity of a range of experimental scenarios in \cref{fig:Omega_gw_experiments} and if $T_\mathrm{rh} \gtrsim 2.1\times 10^2\,\mathrm{GeV}$ the predicted signal is out of reach for even the futuristic proposals. However, for the extremely low reheating temperature of $\sim 100\,\mathrm{MeV}$, the SGWB sourced by structure formation in the matter-dominated post-inflationary era would be detectable by BBO and the Einstein telescope. 

The underlying inflationary scenario in this analysis is an unrealistic toy model from the perspective of current observations. As a consequence, the initial perturbation spectrum may differ from that assumed here. This could change the value of $\mathcal{N}$ at the onset of nonlinear collapse, and modify the form of the resulting SGWB. These differences need not be dramatic but would be more significant in models where the amplitude of the primordial  perturbations rises at very short scales.  

Restricting ourselves to pure N-body simulations, the numerically obtained SGWB misses potential contributions from the formation of inflaton stars and their mergers. Solving the  Schrödinger-Poisson equations requires resolving the de Broglie wavelength, as in the simulations in Refs.~\cite{Musoke2019, Eggemeier2021}, which limits the spatial extent of the box size and for how many $e$-folds after inflation the simulations can be evolved. 
Since running  Schrödinger-Poisson simulations comparable to our N-body simulations in both spatial and temporal extent is an intractable challenge, one could instead approach the problem by computing the gravitational wave signal associated with the formation of inflaton stars and from binary mergers separately. However, due to their relatively small spatial size, one can expect that this will have an effect only at frequencies larger than those covered by the SGWB from our N-body simulations. 

In many cases, inflation is followed by parametric resonance and preheating which  produces its own distinctive gravitational wave background~\cite{Khlebnikov1997_grav, Easther2006_1, Easther2006_2, Garcia-Bellido2007_1, Easther2007, Garcia-Bellido2007_2, Dufaux2007, Dufaux2008, Figueroa2017, Zhou2013, Antusch2016}.  Resonance is not synonymous with thermalization and the post-resonance universe can easily be effectively matter-dominated~\cite{Lozanov:2016hid}. Resonance typically occurs shortly after inflation ends and it leaves the universe highly inhomogeneous. Consequently,  the total duration of the nonlinear phase following resonance could be far longer than in the scenario considered here, substantially increasing the scope for gravitational wave production as a result of gravitationally driven nonlinear dynamics. Conversely, a long matter-dominated phase after resonance would dilute the SGWB produced during resonance itself since the fractional energy density in gravitational waves, $\Omega_{\mathrm{gw}}$, scales as $a^{-1}$ during matter-dominated growth.

\section{Discussion and Conclusions}
\label{sec:conclusions}

Using N-body simulations, we have numerically computed the SGWB originating from the formation of inflaton halos during an epoch of gravitational structure formation following inflation. In the specific example we consider, the inflaton field is evolved from $\mathcal{N}=14$ to $\mathcal{N}=23$ $e$-folds after the end of inflation, at which time complex gravitationally bound structures will have formed. With instantaneous thermalization at the end of the simulation, the universe would reheat to a temperature of $T_\mathrm{rh} = 7.3\times 10^7\,\mathrm{GeV}$ and the resulting gravitational wave spectrum is shown in \cref{fig:Omega_gw_experiments}.  This signal is far below the sensitivity curves of proposed experiments.

Comparing the numerically computed spectrum at $\mathcal{N}=23$ to the corresponding signal from the collapse of density fluctuations into inflaton halos calculated with the quadrupole approximation, we find that other small-scale contributions dominate the SGWB at high frequencies. These include merger events and  tidal interactions among the inflaton halos that occur abundantly on small scales at comparatively early times. We use the approach developed in Ref.~\cite{Jedamzik2010_grav} to extend the spherical collapse approximation by introducing an efficiency parameter $\xi$ for the emission of gravitational waves between halo collapse and reheating, and an additional scale factor contribution $(a_\mathrm{rh}/a_\mathrm{coll})^p$. 

The free parameters $\xi$ and $p$ can be calibrated against the numerical gravitational wave spectrum. This process yields a good fit to the full numerical outcome throughout the simulations and with this result in hand, we can extrapolate the gravitational wave spectrum to smaller reheating temperatures. The present-day amplitude of the spectrum increases and its frequencies decrease as the matter-dominated phase continues. This calculation suggests that the SGWB could be potentially observable by future experiments (see \cref{fig:Omega_gw_experiments}), although only when thermalization occurs at $100\,\mathrm{MeV}$, a value that is probably unrealistically low.  

That said, the analysis here is based on the simplest possible inflationary scenario, namely the purely quadratic potential that is not consistent with cosmological observations. 
The precise shape and the overall amplitude of the computed SGWB will depend on the detailed form of the model and on the amplitude of the  modes that leave the horizon as inflation ends.
To date, there have been no self-consistent analyses of the nonlinear matter-dominated phases for inflationary models that are consistent with astrophysical bounds on the primordial spectra, so this is clearly a very promising  topic for future investigation.

It is well known that resonance  can lead to the production of large density fluctuations which would accelerate the subsequent gravitational formation of structure. A long matter-dominated phase in the post-inflationary universe following resonance will dilute this gravitational wave signal. Conversely, the inhomogeneity generated during resonance will mean that gravitational collapse can begin a few $e$-folds after inflation, significantly enhancing the potential amplitude of the signal studied here. Understanding this tradeoff in realistic scenarios is an important line of enquiry. 

Additional contributions to the SGWB that cannot be captured by our simulations originate from the evaporation of the inflaton halos once reheating takes place~\cite{Jedamzik2010_grav}. Provided that the matter-dominated era of early structure formation continues sufficiently long, it is possible that inflaton halos eventually collapse into a PBH~\cite{Eggemeier2021, Padilla2022, Hidalgo2022}; see also Ref.~\cite{Martin:2019nuw} for another discussion of PBH formation prior to reheating. 
This  will likewise involve the emission  of gravitational waves that leave a further characteristic imprint on the SGWB.    

In summary, the analysis here has quantified a largely unexplored source of a stochastic background of gravitational waves generated in the primordial universe. Its full properties -- along with the detailed nonlinear gravitational dynamics of the post-inflationary matter-dominated phase -- are still unknown. However, our results here make it clear that this phase might have directly observable consequences in the present-day universe. These have the potential to open a window into the earliest moments after the Big Bang and may also provide further incentives for the development of highly sensitive gravitational wave detectors.

\section*{Acknowledgements}
We thank Mona Dentler, Mateja Gosenca, Peter Hayman, Sebastian Hoof, Emily Kendall, Bodo Schwabe, and Yourong (Frank) Wang for useful discussions.
The authors gratefully acknowledge the computing time granted by the Resource Allocation Board and provided on the supercomputer Lise and Emmy at NHR@ZIB and NHR@Göttingen as part of the NHR infrastructure. The calculations for this research were conducted with computing resources under the project nip00052. 
BE acknowledges support from the Deutsche Forschungsgemeinschaft.
RE acknowledges support from the Marsden Fund of the Royal Society of New Zealand. This collaboration was supported by a Julius von Haast Fellowship Award provided by the New Zealand Ministry of Business, Innovation and Employment and administered by the Royal Society of New Zealand.

\bibliography{refs}

\end{document}